\documentclass[reprint,english,aps,prl,twocolumn,amsmath, amssymb, showpacs,showkeys]{revtex4-1}
\usepackage[T1]{fontenc}
\usepackage{subfigure}
\usepackage[latin1]{inputenc}
\usepackage{graphicx}
\usepackage{epstopdf}
\usepackage{epsfig}
\usepackage{prettyref}
\usepackage{babel}
\usepackage{color}
\makeatother

\begin{document}

\title{Observation of multichannel quantum coherent transport and electron-electron interaction in Bi$_2$Te$_3$ single crystal}

\author{Archana Lakhani}
\email{archnalakhani@gmail.com}\affiliation{UGC-DAE Consortium for Scientific Research, University Campus, Khandwa Road, Indore-452001, India}

\author{Devendra Kumar}
\email{deveniit@gmail.com}\affiliation{UGC-DAE Consortium for Scientific Research, University Campus, Khandwa Road, Indore-452001, India}

\begin{abstract}
The bulk of topological insulators is relatively unexplored due to excess contribution of conduction from native defects. Here we investigate the bulk conduction in a Bi$_2$Te$_3$ crystal having reduced defect induced conduction. Our results uncover the presence of three transport regimes which are dominated by thermal activation across bulk band gap, defect state charge conduction, and quantum coherent transport. The low temperature conductance and magnetoconductance reveal the presence of multichannel two dimensional quantum coherent transport in the bulk. The number of channels are of the order of quintuple layers, signifying each quintuple layer as a single transport channel. These transport channels exhibit two dimensional electron-electron interaction effect causing electron dephasing whereas the defect state bulk conduction exhibits three dimensional electron-electron interaction effect which vanishes on enhancement in defect induced charge carriers.
\end{abstract}



\maketitle

\section{Introduction}
The three dimensional (3D) topological insulators are interesting class of materials having an insulating bulk and conducting surface states possessing  spin momentum locked two dimensional Dirac electrons which are immune to backscattering. These surface states are protected by the topology of the bulk band and time reversal symmetry~\cite{Hasan, Qi}. The topological materials possess interesting magnetotransport properties like large linear magnetoresistance~\cite{Kumar1,Wang2,Barua,Shrestha,Qu1}, $\pi$ Berry phase in Shubnikov-de Haas (SdH) oscillations~\cite{Qu1,Yan1,Analytis1,Kumar2}, weak antilocalization (WAL)~\cite{Kim1,Takagaki,Chiu,Wang4,Shekhar}, quantum anomalous Hall effect~\cite{Chang}, quantum Hall effect~\cite{Yoshimi}, Aharonov-Bohm effect~\cite{Peng}, and universal conduction fluctuations~\cite{Li2}. To harness these transport properties, surface dominated conduction with large dephasing length is required which is attempted  by optimizing growth conditions~\cite{Vries,Banerjee1}, charge compensation doping~\cite{Hattori,Ren,Ren1}, and electrostatic gating~\cite{Chiu}. A comprehensive insight of the transport mechanism  in different conduction regimes of topological insulators is needed to understand the fundamental issues like quantum interference effects, phase coherent transport, effect of disorder and electron-electron interactions on topological state~\cite{Bardarson,Liao1}.

The 3D topological insulators like Bi$_2$Se$_3$ consists of quintuple layers of Se1-Bi-Se2-Bi-Se1 separated by the Van der Waal gap. The weak Van der Waal coupling between these quintuple layers results in a highly anisotropic electronic structure~\cite{Eto,Yavorsky}. The presence of vacancies, defects and intersite occupancy contribute to bulk conduction resulting in a metal like behavior. In our recent work~\cite{Kumar3,Tayal} and some of earlier reports~\cite{Cao,Chiatti,Ge}, a bulk quantum Hall effect like feature is observed which depends on the number of quintuple layers in sample. These experimental results demand careful consideration on the understanding of  bulk transport in these materials. In particular, we need to investigate the possibility of two dimensional (2D) transport channels in bulk, their correlation with quintuple layers and their signature in transport properties. Most importantly, the relation between these transport channels and the bulk electrons responsible for metal like behavior in these topological insulators needs to be resolved.

The quantum coherent transport is an important phenomena for investigating the existence of two dimensional electron channels and their coupling. Quantum coherent transport in microflakes and thin films of topological insulators have been extensively utilised for identification of topological surface states~\cite{Kim1,Takagaki,Chiu,Wang4,Shekhar}. In this work, we uncover the underlying physics of bulk conduction in topological insulators by magnetotransport studies on Bi$_2$Te$_3$ crystal having reduced defect state conduction. Our results show the presence of multiple 2D transport channels in the bulk. The number of transport channels are of the order of number of quintuple layers suggesting that each quintuple layer acts as a 2D transport channel. The metal like bulk transport in topological insulators arises from contribution of defect states to conduction which freezes out at low temperatures revealing the clear signature of 2D quantum coherent transport and electron-electron interaction. The  bulk conduction from defect states exhibit 3D electron-electron interaction at low temperatures.

\section{Experimental Details}
\begin{figure}
\begin{centering}
\includegraphics[width=1.0\columnwidth]{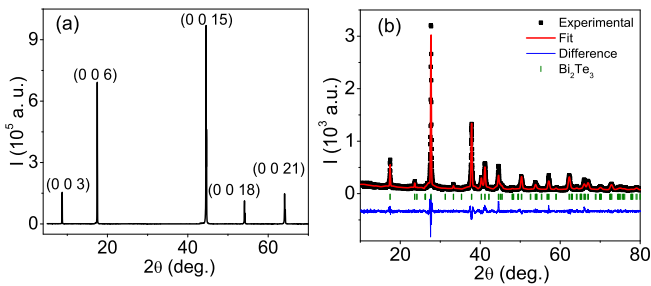}
\par\end{centering}
\caption{(Color Online) $\theta-2\theta$ X-ray diffraction pattern of (a) freshly cleaved Bi$_2$Te$_3$ single crystal and (b) Bi$_2$Te$_3$ powder obtained from crushing of single crystal. The square symbols correspond to the experimental data, the solid red line shows the Rietveld refinement for $R3\bar{m}$ space group, short vertical lines show the Bragg peak positions, and the solid blue line shows the difference between the experimental data and fit.} \label{fig: XRD}
\end{figure}

Single crystal of Bi$_2$Te$_3$ was prepared following our previously reported method on Bi$_2$Se$_3$~\cite{Kumar1,Kumar2} with a slight variation in heating protocol. The initial mixture was prepared using elements of Bismuth and Tellurium having purity of 99.999\%. The structural characterization of the Bi$_2$Te$_3$ crystal was done by X-ray diffraction (XRD) on D8 advanced diffractometer from Brucker using Cu K$\alpha$ radiation ($\lambda$=1.54~\AA). Figure~\ref{fig: XRD}~(a) shows the XRD pattern of the freshly cleaved surface and \ref{fig: XRD}~(b) shows the XRD pattern of the crushed crystal along with its calculated fit using FULLPROF software. The XRD data of the plane surface demonstrates the peaks only corresponding to \{0 0 3\} planes which shows the cleaved surface is perpendicular to the $C_3$ axis.  The Reitveld refinement confirms the $R3\bar{m}$ space group and the lattice parameters obtained are $a$=$b$=4.3877~\AA~ and $c$=30.5122~\AA~ which are in agreement with the reported literature~\cite{Sultana,Cava}. The electrical resistivity and magnetoresistance measurements were performed on two freshly cleaved samples S1 and S2 from the same crystal using 9T PPMS-AC transport measurement system.  The four linear Ohmic contacts were prepared on freshly cleaved thin rectangular shaped piece of Bi$_2$Te$_3$ crystal using silver paste.

\begin{figure}[t]
\begin{centering}
\includegraphics[width=0.9\columnwidth]{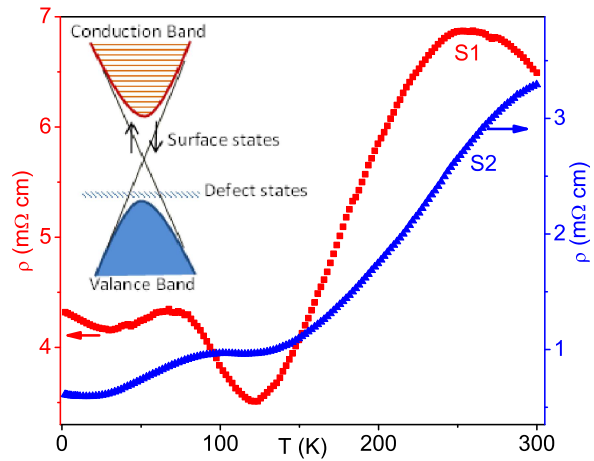}
\par\end{centering}
\caption{(Color Online) Temperature dependence of resistivity  for Bi$_2$Te$_3$ single crystal cleaved samples S1 (red squares) and S2 (blue triangles). Inset shows the schematic band picture used in discussion of results.} \label{fig: RT}
\end{figure}

\begin{figure} [t]
\begin{centering}
\includegraphics[width=0.9\columnwidth]{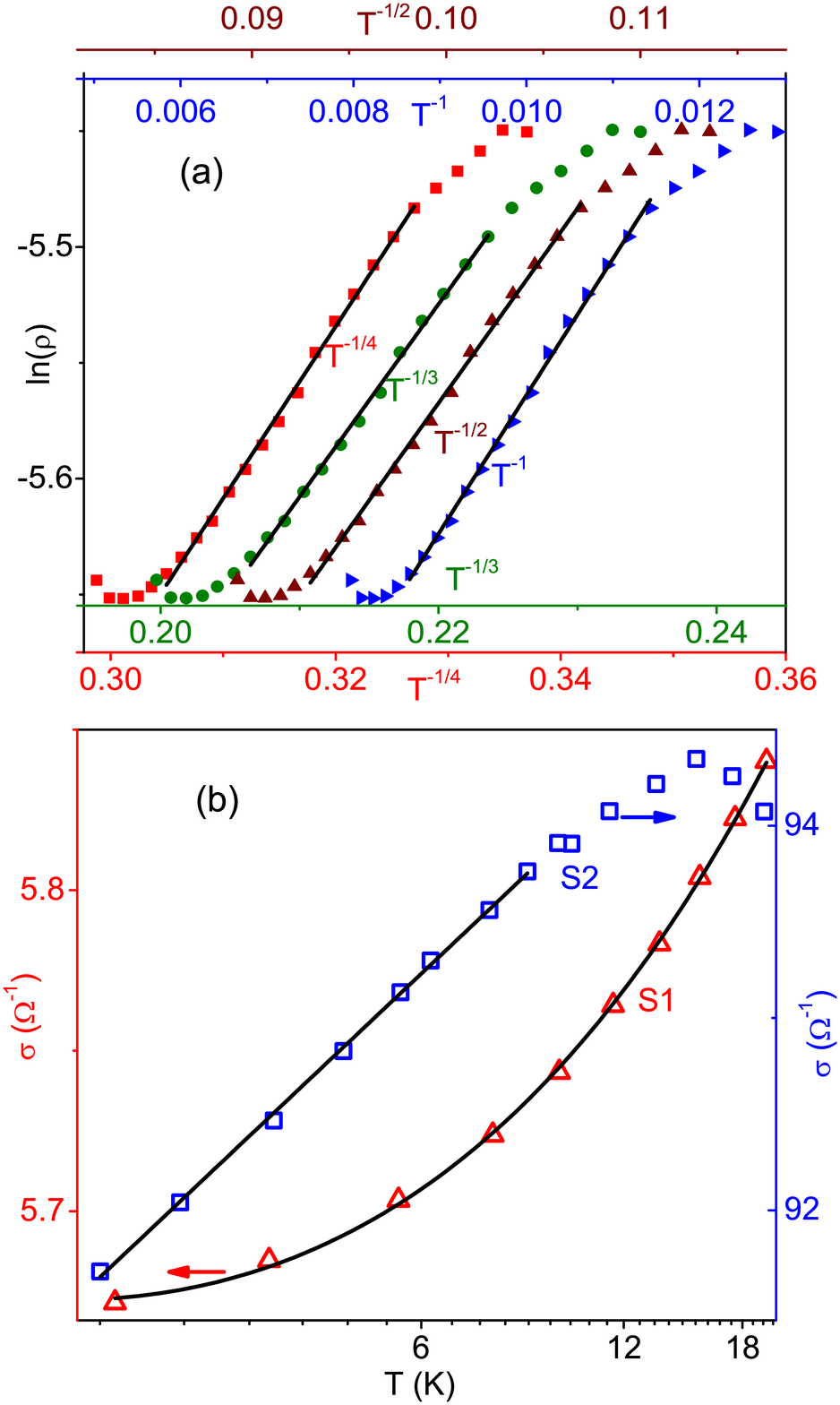}
\par\end{centering}
\caption{(Color Online) (a) Plot of ln($\rho$) versus $T^{-1}$, $T^{-1/2}$, $T^{-1/3}$, and $T^{-1/4}$ for sample S1. Solid lines are the least square fitting of straight line. (b) Conductivity versus temperature for sample S1 (red triangles) and S2 (blue squares). Solid lines are the least square fitting of the combination of 2D EEI, WAL and 3D EEI theory to S1 and the combination of 2D EEI and WAL theory to S2.} \label{fig: RTfit}
\end{figure}

\section{Results and Discussion}
Figure~\ref{fig: RT} shows the temperature variation of resistivity on cooling in zero magnetic field for samples S1 and S2. For S1, the resistivity initially increases on lowering the temperature and crosses over to a typical metal like behavior at $\sim$250~K. On further lowering the temperature, below 120~K, the resistivity starts increasing and attains a  maximum around 75~K, then decreases and attain a minimum around 30~K with an upward trend below this temperature. Such multiple changes in the slope of temperature dependent resistivity suggest the interplay of multiple transport mechanism in our crystal.
This also indicates that Fermi level in our crystal does not lie deep in the valance band as in the case for most stoichiometric Bi$_2$Te$_3$ crystals~\cite{Cava}. The insulating behaviour of resistivity in the temperature range 250~K-300~K suggests that transport is dominated by the thermal excitation of electrons from valance to conduction band. The number of thermally excited charge carriers decrease on lowering the temperature and the change in resistivity behavior at 250~K indicates the transport is dominated by the holes  in valance band arising due to transfer of electrons to shallow defect states near the top of valance band. See the schematic band picture shown in the inset of Figure~\ref{fig: RT}. The increase in resistivity below 120~K could be due to frozen out effect of charge carriers from bulk valance to shallow defect band or from the disorder induced localization effects in the bulk~\cite{He2,Hattori,Shekhar,Ren,Ren1}.

The resistivity can be fitted with $\rho$=$\rho_0$exp$(\Delta/k_BT)$ for frozen out effect or with $\rho$=$\rho_0$exp$(T_0/T)^{1/\nu+1}$ for hopping between localized states where $\nu$=1 for one-dimensional or Efros-Shklovskii, $\nu$=2 for two-dimensional, and $\nu$=3 for three-dimensional variable range hopping~\cite{Mott}. Figure \ref{fig: RTfit}~(a) shows the ln$\rho$ versus $T^{-1}$, $T^{-1/2}$, $T^{-1/3}$, and $T^{-1/4}$ plot for resistivity in the temperature range of 125 to 77~K. The best linear fit is obtained for $T^{-1}$ plot giving activation energy $\Delta$=5.1(1) meV.
The almost saturated resistivity below 75~K signals the insignificance of bulk contribution in overall conduction below this temperature, therefore the conductance of the system in this temperature range should ideally be $\sim$ $e^2/h$ which is the conductance of one surface channel of the topological insulator. The conductance of our system at around 70~K is 5.67~$\Omega^{-1}$ ($\approx$ 1.46$\times10^5e^2/h$) which suggests the presence of multiple transport channels in our crystal. Sample S2 also exhibits similar conduction regimes, but the metal like conduction starts above 300~K indicating a higher contribution of defect induced charge carriers. The relative flat behavior of resistivity between 115~K-100~K suggests that the freezing/localization of defect induced charge carriers at low temperature is relatively weak in comparison to S1.

\begin{figure*}[]
\begin{centering}
\includegraphics[width=1.0\textwidth,height=9.9cm]{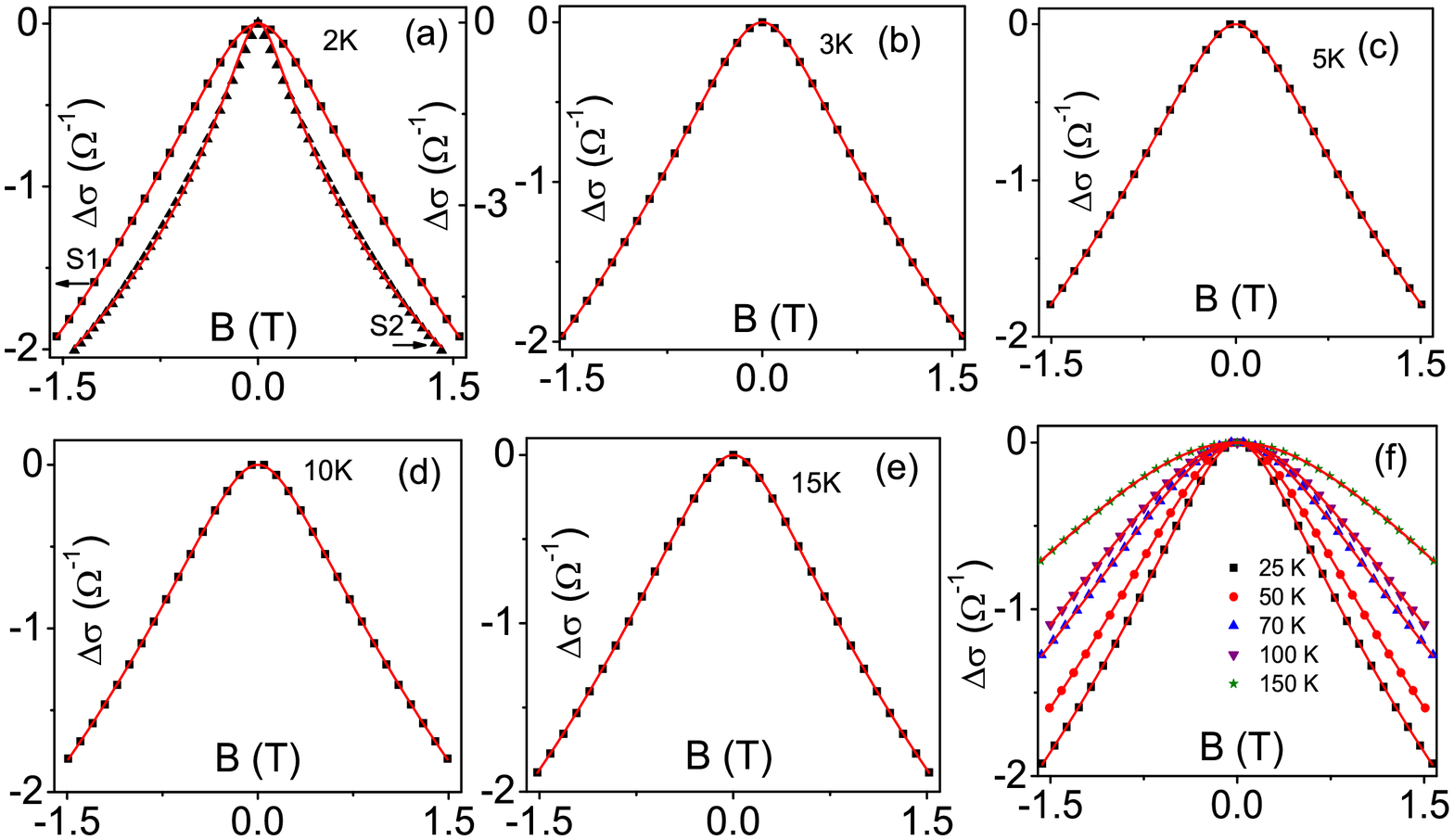}
\par\end{centering}
\caption{(Color Online) (a) Low field magnetoconductance versus field for sample S1 (squares) and S2 (triangles) at 2~K. Magnetoconductance of S1 at (b) 3~K, (c) 5~K, (d) 10~K, (e) 15~K and (f) 25~K, 50~K, 70~K, and 100~K. Solid lines are the best fit of HLN equation.  }
\label{fig: HLN}
\end{figure*}


The decrease in resistivity below 70~K and the upward trend below 30~K in S1 is the effect of weak antilocalization and electron-electron interaction (EEI) corrections to resistivity respectively. In quantum transport regime, the quantum interference between the time reversed electron waves causes weak localization (WL) or weak anti-localization (WAL) correction to conductivity. WL is realized in limit of weak spin orbit scattering while WAL is observed in limit of strong spin orbit scattering or from $\pi$ Berry phase in topological surface states~\cite{Hikami,Lee,McCann}. The weak  localization correction to conductivity $\Delta\sigma_{}$ is negative for WL and positive for WAL. For 3D, the WAL correction to conductivity $\Delta\sigma_{\text{3DWAL}}\propto -T^{p/2}$ where $p$ is the index which determines the temperature dependence of inelastic scattering time causing electron dephasing and depends on the scattering mechanism, dimensionality etc. For 2D, the WAL correction to conductivity is given as
\begin{equation}
\Delta\sigma_{\text{2DWAL}}=-\alpha\frac{p}{2}\frac{e^2}{\hbar\pi^2}\text{ln}\frac{T}{T_0},\label{eq:2DWAL}
\end{equation}
where $\alpha$= 1/2 for single channel WAL and $T_0$ is the temperature above which WAL correction vanishes. $p\geq$2(1) for 3D(2D) electron phonon scattering, 3/2 for 3D EEI, and 1 for 2D EEI~\cite{Takagaki,Wang4, Yuan1,Lin,Rammer,Lawrence}. The dephasing mechanism is generally dominated by electron-phonon scattering in 3D weak disorder systems, while for 2D systems, the Nyquist electron-electron scattering is the dominant scattering process at low temperatures.

The electron-electron interaction (EEI) effect in a disordered system also gives rise to corrections in conductivity at low temperatures. For 3D, the EEI correction to temperature dependence of conductivity is~\cite{Lee}
\begin{equation}
\Delta\sigma_{\text{3DEEI}}=\frac{e^2}{2\hbar\pi^2}\frac{1.3}{\sqrt{2}}(\frac{2}{3}-\frac{3}{4}\tilde{F_\sigma})\sqrt{T/D},\label{eq:2EEI}
\end{equation}
and for 2D the EEI correction is
\begin{equation}
\Delta\sigma_{\text{2DEEI}}=\frac{e^2}{2\hbar\pi^2}(1-\frac{3}{4}\tilde{F_\sigma})\text{ln}\frac{T}{T_0},\label{eq:2EEI}
\end{equation}
where the coefficient $\tilde{F_\sigma}$ is the electron screening factor, $T_0$ is the characteristic temperature for EEI effect and $D$ is the diffusion constant. The conductance $\sigma(T)$ of S1 and S2 below 20~K is shown in Figure~\ref{fig: RTfit}~(b) and is fitted with  a combination of 2D WAL and EEI ($\Delta\sigma$=$a\text{ln}T$), 3D WAL and EEI ($\Delta\sigma$=$bT^{p/2}+c\sqrt{T}$), 3D WAL and 2D EEI/WAL ($\Delta\sigma$=$a\text{ln}T+bT^{p/2}$), and 2D WAL/EEI and 3D EEI ($\Delta\sigma$=$a\text{ln}T+c\sqrt{T}$ ) models. The coefficient of WAL term should be negative and the coefficient of EEI term should be positive. The best fit with physically significant parameters is obtained by the 2D WAL/EEI and 3D EEI combination, suggesting the presence of 2D WAL, EEI and 3D EEI in S1; whereas  $\sigma(T)$ of S2 fits well with 2D WAL and EEI combination indicating the presence of only 2D WAL and EEI in S2. The strength of 3D electron-electron interaction $\propto n^{-1/3}$, where $n$ is the carrier density. Since S1 has lower defect induced charge carriers which is further reduced by stronger freezing/localization effects at low temperatures in comparison to S2, the observation of 3D EEI in S1 and its absence in S2 suggests that the 3D EEI effect in S1 arises due to defect induced charge carriers.

\begin{figure} [t]
\begin{centering}
\includegraphics[width=0.8\columnwidth]{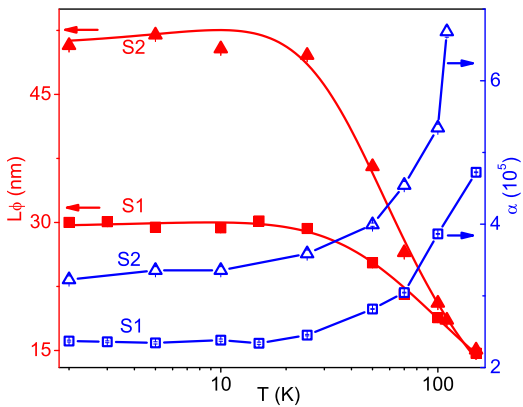}
\par\end{centering}
\caption{(Color Online) Temperature dependence of dephasing length $L_{\phi}$ and $\alpha$ for sample S1 ($L_{\phi}$ solid squares, $\alpha$ open squares) and S2 ($L_{\phi}$ solid triangles, $\alpha$ open triangles). Solid lines are the best fit of Equation~\ref{eq:Depheq}.} \label{fig: Deph}
\end{figure}

The low field magnetoconductance for WAL effect is expected to follow Kawabata~\cite{Kawabata} equation for 3D WAL and Hikami-Larkin-Nagaoka (HLN)~\cite{Hikami} equation for 2D WAL. The least square fitting of low field magnetoconductivity of S1 with Kawabata equation gives prefactor $\alpha$=480 which is theoretically 1 and is observed to lie between $\sim$ 0.3-0.5 in the experiments~\cite{Hattori,Dynes}. The unphysical value of  $\alpha$ rules out the possibility of 3D WAL, and as seen by the temperature dependence of conductivity, a 2D WAL model is more appropriate for describing the magnetoconductance of our system. The low field magnetoconductance is fitted  with Hikami-Larkin-Nagaoka (HLN) equation~\cite{Hikami}
 \begin{equation}
\Delta\sigma(B)=-\alpha\frac{e^2}{2\hbar\pi^2}\left[\Psi\left(\frac{1}{2}+\frac{\hbar}{4eL^2_{\phi}B}\right)-\text{ln}\left(\frac{\hbar}{4eL^2_{\phi}}\right)\right],\label{eq:2EEI}
\end{equation}
where $\alpha$= 1/2 for single channel WAL,  $\alpha$= $\frac{1}{2}n_c$ for $n_c$ independent 2D transport channels with same dephasing time, $\Psi$ is the digamma function and $L_{\phi}$ is the dephasing length which is the average length an electron can maintain its phase. For single topological surface state $\alpha$=0.5  and for two identically decoupled topological surface states separated by an insulating bulk $\alpha$=1. The coupling between the metallic bulk state and the two surface states gives $\alpha$ =0.5~\cite{Kim1,Liao}. Experimentally $\alpha$=0.3-1.1 is observed in thin films depending on the coupling between bulk and surface states~\cite{Wang4,Takagaki,Lu2,Jing1} and for single crystals $\alpha\approx$3 has been reported for insulating Bi$_2$Te$_2$Se~\cite{Shekhar}  with resistance of 1k$\Omega$ at 2~K  while $\alpha\sim$ 10$^5$ for bulk crystals with metallic transport~\cite{Barua1,Shrestha}. Figure~\ref{fig: HLN} shows the symmetrized low field magnetoconductance of S1 and S2 along with the least square fitting of HLN equation at different temperatures. The magnetoconductance data fits well with the HLN equation upto $\sim$ 150~K and the values of fitting parameters $L_{\phi}$ and $\alpha$ are shown in Figure~\ref{fig: Deph}.

The dimensionality of WAL effect is determined by the relation between dephasing length $L_{\phi}$ and sample thickness $t$. For $L_{\phi}$<$t$  WAL effect is 3D while for $L_{\phi}$>$t$ WAL effect is 2D. For S1 $t$=0.09~mm, $L_{\phi}$=30~nm and for S2 $t$=0.236~mm, $L_{\phi}$=51~nm at 2K. Both of our samples have $L_{\phi}$<$t$ suggesting the possibility of 3D WAL effect whereas the temperature dependence of conductance and the HLN fitting of magnetoconductance show a 2D WAL effect. In Bi$_2$Te$_3$, the two dimensional quintuple layers of thickness $\approx$1~nm~\cite{Teweldebrhan} are weakly coupled along the $c$ axis and it seems that these quintuple layers act as 2D transport channels. A single quintuple layer have been observed to act as an independent transport channel in Qunatum Hall effect scaling of Bi$_2$Se$_3$~\cite{Kumar3,Cao} and Fe-doped Bi$_2$Se$_3$~\cite{Ge} single crystals while two quintuple layers act as a single transport channel in case of Bi$_2$Se$_3$ microflakes~\cite{Chiatti}. The thickness of a quintuple layer is much less than the dephasing length in our samples ($L_{\phi}$=30~nm for S1 and $L_{\phi}$=51~nm for S2) suggesting that 2D WAL effect can arise if each (or a couple of) quintuple layer  acts like a 2D transport channel. For S1, $t$=0.09~mm allow $\approx$ 9$\times$10$^4$ quintuple layers and if each quintuple layer acts as an independent transport channel, we expect around 9$\times$10$^4$ 2D transport channels participating in the WAL effect. At 2~K, the HLN fitting of magnetoconductance gives $\alpha$=2.37(1)$\times$10$^5$ suggesting the presence of 4.74$\times$10$^5$ independent 2D transport channels.  For sample S2, $t$=0.236~mm allow $\approx$ 2.36$\times$10$^5$ 2D transport channels while HLN fitting of magnetoconductance gives $\alpha$=3.22$\times$10$^5$ suggesting the participation of 6.44 $\times$10$^5$ independent 2D transport channels in the WAL. The number of 2D transport channels obtained from HLN fitting and from quintuple layers are of the same order and both increase with sample thickness, indicating that our crystal has multiple 2D transport channels and each quintuple layer in the sample acts like a 2D transport channel. The relatively large number of transport channels obtained from HLN fitting could be due to coupling of 2D transport channels with the residual defect induced bulk conduction~\cite{Shekhar}.

Figure~\ref{fig: Deph} shows the temperature dependence of electron dephasing length $L_{\phi}$ for S1 and S2. $L_{\phi}$ remains constant upto $\sim$20~K and decreases on further increasing the temperature due to enhancement in inelastic scattering. The temperature dependence of $L_{\phi}$ can be explained with~\cite{Wang4,Shrestha}
\begin{equation}
\frac{1}{L^2_{\phi}(T)}=\frac{1}{L^2_{\phi}(0)}+A_{ee}T^{p1}+A_{ep}T^{p2},\label{eq:Depheq}
\end{equation}
where $L_{\phi}(0)$ is the zero temperature dephasing length which depends on various factors including sample geometry and defects,  $A_{ee}T^{p1}$ and $A_{ep}T^{p2}$ are the contributions of electron-electron (e-e) and electron-phonon (e-ph) scattering respectively.  $p1$ and $p2$ depend on the temperature dependence of inelastic scattering time with $p1$=1 for 2D e-e scattering and $p2\geq$1 for 2D  e-ph scattering. The best fitting of Equation~\ref{eq:Depheq} to  $L_{\phi}(T)$ is obtained for $A_{ee}$=-1.7$\times$10$^{-5}$, $A_{ep}$=4.6(5)$\times$10$^{-6}$, $p1$=1, $p2$=1.4 for S1 and $A_{ee}$=-7(2)$\times$10$^{-6}$, $A_{ep}$=4.3(4)$\times$10$^{-7}$, $p1$=1 and $p2$=1.9 for S2  suggesting that 2D electron-electron scattering is the dominant dephasing mechanism along with the 2D electron-phonon scattering. Further, the two dimensional nature of the dephasing mechanism reiterates the fact that WAL effect is emerging from 2D transport channels.

 \section{Conclusion}
 In conclusion, we have studied temperature dependent conductance and low field magnetoconductance of  Bi$_2$Te$_3$ single crystal. Our results demonstrate the presence of three charge transport regimes viz. thermal excitation across the bulk bands, defect induced conduction and quantum coherent transport dominated at high, intermediate and low temperature regions respectively. The comprehensive analysis of quantum coherent transport in the bulk single crystal reveals the presence of multiple 2D transport channels. The number of channels and quintuple layers are of same order in both the samples studied and are directly related to the thickness of the sample, revealing each quintuple layer acting as a transport channel. Further, 2D transport channels exhibit the 2D EEI effect whereas defect induced conduction reveals the presence of 3D EEI effect which vanishes on the enhancement of defect state charge conduction.

\section{Acknowledgements}
We thank M. Gupta and L. Bahera for XRD measurements, S. Rana and R. P. Jena for help in sample preparation. V. Ganesan and A. K. Sinha are acknowledged for support and encouragement.


\begin{thebibliography}{10}
\bibitem{Hasan} M. Z. Hasan  and C. L. Kane, Rev. Mod. Phys. \textbf{82}, 3045 (2010).

\bibitem{Qi} X. -L. Qi  and S. -C. Zhang, Rev. Mod. Phys. \textbf{83}, 1057 (2011).

\bibitem{Kumar1} D. Kumar and A. Lakhani, Phys. Status Solidi RRL \textbf{12}, 1800088 (2018).


\bibitem{Wang2} X. Wang, Y. Du, S. Dou, and C. Zhang, Phys. Rev. Lett. \textbf{108}, 266806 (2012).

\bibitem{Barua} S. Barua, K. P. Rajeev, and A. K. Gupta, J. Phys.: Condens. Matter \textbf{27}, 015601 (2015).

\bibitem{Shrestha} K. Shrestha, M. Chou, D. Graf, H. D. Yang, B. Lorenz, and C. W. Chu, Phys. Rev. B \textbf{95}, 195113 (2017).


\bibitem{Qu1} D.-X. Qu,  Y. S. Hor, J. Xiong, R. J. Cava, and N. P. Ong, Science \textbf{329}, 821 (2010).

\bibitem{Yan1} Y. Yan, Z.-M. Liao, Y.-B. Zhou, H.-C. Wu, Y.-Q. Bie, J.-J. Chen, J. Meng, X.-S. Wu, and D.-P. Yu, Sci. Rep. \textbf{3}, 1264 (2013).

\bibitem{Analytis1} J. G. Analytis, R. D. McDonald,	 S. C. Riggs,  J. -H. Chu,  G. S. Boebinger, and I. R. Fisher, Nat. Phys. \textbf{6}, 960 (2010).

\bibitem{Kumar2} D. Kumar and A. Lakhani, Phys. Status Solidi RRL \textbf{9}, 636 (2015).

\bibitem{Kim1} Y. S. Kim, M. Brahlek, N. Bansal, E. Edrey, G. A. Kapilevich, K. Iida, M. Tanimura, Y. Horibe, S. -W. Cheong  and S. Oh, Phys. Rev. B \textbf{84}, 073109 (2011).


\bibitem{Takagaki} Y. Takagaki, B. Jenichen, U. Jahn, M. Ramsteiner, and K. -J. Friedland, Phys. Rev. B \textbf{85}, 115314 (2012).

\bibitem{Chiu} S. -P. Chiu and J. -J. Lin, Phys. Rev. B \textbf{87}, 035122 (2013).

\bibitem{Wang4} W. J. Wang, K. H. Gao, and Z. Q. Li, Sci. Rep. \textbf{6}, 25291 (2016).

\bibitem{Shekhar} C. Shekhar, C. E. ViolBarbosa, B. Yan, S. Ouardi, W. Schnelle, G. H. Fecher, and C. Felser, Phys. Rev. B \textbf{90}, 165140 (2014).

\bibitem{Chang} C.-Z. Chang, J. Zhang, X. Feng, J. Shen, Z. Zhang, M. Guo, K. Li, Y. Ou, P. Wei, L.-L. Wang, Z.-Q. Ji, Y. Feng, S. Ji, X. Chen, J. Jia, X. Dai, Z. Fang, S.-C. Zhang, K. He, Y. Wang, L. Lu, X.-C. Ma, Q.-K. Xue, Science \textbf{340}, 167 (2013).

\bibitem{Yoshimi} R. Yoshimi, A. Tsukazaki, Y. Kozuka, J. Falson, K. S. Takahashi, J. G. Checkelsky, N. Nagaosa, M. Kawasaki, and Y. Tokura, Nat. Commun. \textbf{6}, 6627 (2015).

\bibitem{Peng} H. Peng, K. Lai, D. Kong, S. Meister, Y. Chen, X.-L. Qi, S.-C. Zhang, Z.-X. Shen, and Y. Cu, Nat. Mater. \textbf{9}, 225 (2010).

\bibitem{Li2} Z. Li, T. Chen, H. Pan, F. Song, B. Wang, J. Han, Y. Qin, X. Wang, R. Zhang, J. Wan, D. Xing, and G. Wang, Sci. Rep. \textbf{2}, 595 (2012).

\bibitem{Vries} E. K. de Vries, S. Pezzini, M. J. Meijer, N. Koirala, M. Salehi, J. Moon, S. Oh, S. Wiedmann, and T. Banerjee, Phys. Rev. B \textbf{96}, 045433 (2017).

\bibitem{Banerjee1} A. Banerjee, R. Ganesan, and P. S. A. Kumar, Appl. Phys. Lett. \textbf{113}, 072105 (2018).

\bibitem{Hattori} Y. Hattori, Y. Tokumoto, and K. Edagawa, Phys. Rev. Materials \textbf{1}, 074201 (2017).

\bibitem{Ren} Z. Ren, A. A. Taskin, S. Sasaki, K. Segawa, and Y. Ando, Phys. Rev. B \textbf{82}, 241306(R) 2010. 

\bibitem{Ren1} Z. Ren, A. A. Taskin, S. Sasaki, K. Segawa, and Y. Ando, Phys. Rev. B \textbf{84}, 165311 (2011).


\bibitem{Bardarson} J. H. Bardarson, and J. E. Moore, Rep. Prog. Phys. \textbf{76}, 056501 (2013).

\bibitem{Liao1} J. Liao, Y. Ou, X. Feng, S. Yang, C. Lin, W. Yang, K. Wu, K. He, X. Ma, Q.-K. Xue, and Y. Li, Phys. Rev. Lett. \textbf{114}, 216601 (2015).

\bibitem{Eto} K. Eto, Z. Ren, A. A. Taskin, K. Segawa, and Y. Ando, Phys. Rev. B \textbf{81}, 195309 (2010).

\bibitem{Yavorsky} B. Y. Yavorsky, N. F. Hinsche, I. Mertig, and P. Zahn, Phys. Rev. B \textbf{84}, 165208 (2011).

\bibitem{Kumar3} D. Kumar and A. Lakhani, Mater. Res. Bull. \textbf{88}, 127 (2017).

\bibitem{Tayal} A. Tayal, D. Kumar, and A. Lakhani, J. Phys.: Condens. Matter \textbf{29}, 445704 (2017).

\bibitem{Cao} H. Cao, J. Tian, I. Miotkowski, T. Shen, J. Hu, S. Qiao, Y. P. Chen, Phys. Rev. Lett. \textbf{108}, 216803 (2012).

\bibitem{Chiatti} O. Chiatti, C. Riha, D. Lawrenz, M. Busch, S. Dusari, J. S.-Barriga, A. Mogilatenko, L. V. Yashina, S. Valencia, A. A. \"Unal, O. Rader, and S. F. Fischer, Sci. Rep. \textbf{6}, 27483 (2016).

\bibitem{Ge} J. Ge, T. Chen, M. Gao, X. Wang, X. Pan, M. Tang, B. Zhao, J. Du, F. Song, Y. Xu, R. Zhang, Sol. State Commun. \textbf{211}, 29 (2015).

\bibitem{Sultana} R. Sultana, P. Neha, R. Goyal, S. Patnaik, V. P. S. Awana, J. Magn. Mag. Mater. \textbf{428}, 213 (2017).

\bibitem{Cava} R. J. Cava, H. Ji, M. K. Fuccillo, Q. D. Gibson, and Y. S. Hor, J. Mater. Chem. C \textbf{1}, 3176 (2013).

\bibitem{He2} L. He, F. Xiu, X. Yu, M. Teague, Wanjun, Jiang, Y. Fan, X. Kou, M. Lang, Y. Wang, G. Huang, N.-C. Yeh, and K. L. Wang, Nano Lett. \textbf{12}, 1486 (2012).

\bibitem{Mott} N. F. Mott, \emph{Metal-Insulator Transitions} (Tayor and Francis, London, 1990) 2nd ed.; B. I. Shklovskii and B. Z. Spivak, in \emph{Hopping Transport in Solids}, edited by M. Pollak and B. I. Shklovskii (North-Holland, Amsterdam, 1991).

\bibitem{Hikami} S. Hikami, A. I. Larkin, and Y. Nagaoka, Prog. Theor. Phys. 63, 707 (1980).

\bibitem{Lee} P. A. Lee and T. V. Ramakrishnan, Rev. Mod. Phys. \textbf{57}, 287 (1985).

\bibitem{McCann} E. McCann, K. Kechedzhi, V. I. Fal'ko, H. Suzuura, T. Ando, and B. L. Altshuler, Phys. Rev. Lett. 97, 146805 (2006).

\bibitem{Yuan1} C.-Yuan Wu, B.-T. Lin, Y.-J. Zhang, Z.-Q. Li, and J.-J. Lin, Phys. Rev. B \textbf{85}, 104204 (2012)

\bibitem{Lin} J. J. Lin, and J. P. Bird, J. Phys.: Condens. Matter \textbf{14}, R501 (2002).

\bibitem{Rammer} J. Rammer and A. Schmid,  Phys. Rev. B 34, 1352 (1986).  

\bibitem{Lawrence} W. E. Lawrence and A. B. Meador, Phys. Rev. B \textbf{18}, 1154 (1978). 

\bibitem{Kawabata} A. Kawabata, J. Phys. Soc. Jpn. \textbf{49}, 628 (1980)

\bibitem{Dynes} R. C. Dynes, T. H. Geballe, G. W. Hull, Jr., and J. P. Garno, Phys. Rev. B \textbf{27}, 5188 (1983).

\bibitem{Liao} J. Liao, Y. Ou, H. Liu, K. He, X. Ma, Q.-K. Xue, and Y. Li, Nat. Commun. \textbf{8}, 16071 (2017).

\bibitem{Lu2} H.-Z. Lu and S.-Q. Shen, Phys. Rev. B \textbf{84}, 125138 (2011).

\bibitem{Jing1} Y. Jing, S. Huang, K. Zhang, J. Wu, Y. Guo, H. Peng, Z. Liub, and H. Q. Xu, Nanoscale \textbf{8}, 1879 (2016).

\bibitem{Barua1} S. Barua and K. P. Rajeev, Sol. State Commun. \textbf{248}, 68 (2016).

\bibitem{Teweldebrhan} D. Teweldebrhan, V. Goyal, M. Rahman, and A. A. Balandin, Appl. Phys. Lett. \textbf{96}, 053107 (2010).





\end{thebibliography}
\end{document}